\documentclass[useAMS,usenatbib,usegraphicx]{mn2e}
\usepackage{multirow}
\usepackage{url}
\usepackage{rotating}

\title[XMM-Newton observations of unidentified INTEGRAL/IBIS sources] 
{\emph{XMM-Newton} observations of unidentified \emph{INTEGRAL}/IBIS sources}

\author[A. Malizia]
{A. Malizia,$^1$\thanks{E-mail address: \texttt{malizia@iasfbo.inaf.it}}L. Bassani,$^1$ V. Sguera,$^1$
J.~B.  Stephen,$^1$ A. Bazzano,$^2$ M. Fiocchi,$^2$ \newauthor and A.~J. Bird$^3$ \\
$^1$ INAF/IASF-Bologna, Via P. Gobetti 101, I-40129 Bologna, Italy \\
$^2$ INAF/IASF-Roma, Via Fosso del Cavaliere 100, I-00133, Roma, Italy \\
$^3$ School of Physics and Astronomy, University of Southampton,       SO17 1BJ, Southampton, UK }

\begin{document}         
\date{Accepted .... Received ...; in original form ...}

\pagerange{\pageref{firstpage}--\pageref{lastpage}} \pubyear{2010}
\maketitle 

\label{firstpage}

\begin{abstract}
About 30\% of the sources in the 4th \emph{INTEGRAL-IBIS} catalogue are unidentified in that they lack an  optical counterpart. To be able to classify them, X-ray observations are of crucial importance as they can place tighter constraints on the high energy error box, which is usually of the order of a few arcminutes, and allow their broad band spectrum to be studied. To this aim we have cross-correlated the list of all unidentified 
IBIS sources in the fourth  catalogue with the archive of all \emph{XMM-Newton} pointings, finding a set of  6
objects with archival data. For 1 of them, IGR J17331--2406, no X-ray source is detected by XMM inside  the IBIS error box, most likely due to the fact 
that it is  a transient object. In the  case of IGR J17445-2747 two possible X-ray counterparts are found inside the IBIS error box:
one is very weak  while the other is bright but only detected once. In each of the remaining 4 cases: IGR J155359-5750, AX J1739.3-2923, AX J1740.2-2903 
and IGR J18538-0102,  we find instead a convincing association
for which we provide an improved X-ray position and information on the optical/infrared counterpart.
We also performed a detailed analysis of their XMM-IBIS spectra and on the basis of all information acquired we
 suggest that IGR J155359-5750 is an AGN of intermediate type, AX J1739.3-2923 and AX J1740.2-2903 are High Mass X-ray Binary  systems,
IGR J17331--2406 and IGR J17445-2747 are Galactic transient sources and IGR J18538-0102 could be a background AGN.
\end{abstract}

\begin{keywords}
catalogues -- surveys -- gamma-rays: observations -- X-rays: observations.
\end{keywords}

\begin{table*}
\begin{center}
\centering
\caption{Log of the \emph{XMM-Newton} and IBIS observations used in this paper.}
\label{tab1}
\begin{tabular}{lcrrrrc}
\hline
\hline
IBIS source                        & XMM Obs. Date    & Expo MOS1 & Expo MOS2 & Expo pn  & IBIS Expo & Var$^{a}$ \\
                                             &                       &   (sec)               & (sec)                   & (sec)                       &     ksec & \\
\hline
\hline
IGR J15359-5750           & 2006-08-15   & 16395          &                       & 11470     & 2839   &  \\
IGR J17331-2406           & 2006-04-02   & 10566          & 10809           &  8937      & 6962   &   Y \\                  
AX J1739.3-2923            & 2005-08-29  &  16899          & 17299          & 14035     & 8047   &   \\
AX J1740.2-2903            & 2005-09-29  &  7568            & 7465            &                 & 8488    &   \\ 
IGR J17445-2747           & 2006-04-04   &  11802         & 11666          & 10149     & 7295    &  YY\\
IGR J18538-0102           & 2004-10-08  &   8058           & 8076             & 5586       & 2955     &  \\     
\hline
 \hline
\end{tabular}
\end{center}
$^{a}$ Variability indicator as in Bird et al. (2010)
\end{table*}

\section{Introduction} 
The fourth and most recent IBIS/ISGRI survey (Bird et al 2010) lists 723 hard X-ray sources  of which
30\% are unidentified, i.e. lack an  optical counterpart. Most of these sources 
are also poorly studied at energies below 10 keV, whereas X--ray
observations can be of crucial importance to unveil their nature and class. POsitional location with
arcsecond accuracy,  which can only be  obtained using the  capabilities of current X-ray telescopes,
is necessary  to pinpoint and classify the optical counterpart of these hard X-ray  emitters.
Furthermore, information in the X-ray band, which is often lacking, is necessary to characterize 
these sources in terms of spectral shape, flux, absorption and variability. 
Since the first IBIS survey, our group has conducted a comprehensive programme in the analysis of X-ray  follow up observations
of unidentified IBIS sources, focusing recently on the 4th IBIS survey objects (Landi et al. 2010, Fiocchi et al. 2010).
With this in mind, we have cross-correlated the list of the still unidentified IBIS sources included in 
the fourth  catalogue with the archive of all \emph{XMM-Newton} pointings, finding a set of  6
objects with archival data.
For one of them no X-ray source is detected by XMM inside  the IBIS error box, most likely due to the fact 
that it is  a variable and/or transient object, while in another case two possible X-ray counterparts are found inside the IBIS error box:
one is very weak  and the other is bright but only detected once. In the remaining 4 cases, we find instead a convincing association
for which we provide the improved X-ray position and information on the optical/infrared counterpart.
We have also performed a detailed analysis of the XMM data and in combination with IBIS spectra  
we have been able to study the broad band properties of each source. All together, these data allow us to investigate the possible 
nature of these hard X-ray sources and to suggest that 1 is mostly likely an AGN, 2 are probably Galactic binaries, 2 are Galactic transient objects and one
could be associated either to a compact object in a supernova remnant or to a background AGN.

\section{Data reduction and analysis}

For the sources in our  sample, we use X-ray data acquired with the three X-ray CCD  cameras (MOS 1, 2  and pn) comprising
the EPIC instrument on-board the  \emph{XMM-Newton} spacecraft (Struder et al. 2001).
The EPIC cameras  offer the possibility to perform sensitive imaging observations over the telescope's field of view (FOV) of 30 arcmin 
and in the energy range from 0.15 to 15 keV with moderate spectral (E/$\Delta E \sim 20-50$) 
and angular resolution (PSF, 6 arcsec FWHM) and 
are therefore  ideal for our objectives  of improving the source position and studying broad band X-ray spectra.

\emph{XMM} data were reprocessed using the  Standard Analysis Software (SAS) version 9.0.0
employing the latest available calibration files.
Only patterns corresponding to single and double events (PATTERN$\leq$4) were taken
into account and the standard selection filter FLAG=0 was applied.
The observations were filtered for periods of high background and the resulting net exposures for each source and each camera
are  reported in table 1 which also lists the XMM observation date.
For each observation, we analysed the \emph{XMM-Newton} EPIC (MOS plus pn) images to search for X-ray sources 
which fall inside the IBIS error box  and are therefore likely IBIS counterparts. Next we obtained X-ray spectra in the 0.5-12 keV band
of the likely associated source.
Source counts were extracted from circular regions of radius 20$^{\prime\prime}$ centered on the source;
background spectra were extracted from circular regions close to the
source or from source-free regions of 80$^{\prime\prime}$ radius. The
ancillary response matrices (ARFs) and the detector response matrices
(RMFs) were generated using the \emph{XMM} SAS tasks \emph{arfgen} and
\emph{rmfgen} while the spectral channels were rebinned in order to achieve a
minimum of 20 counts in each bin.

The INTEGRAL data reported here consist of all pointings performed  (Winkler et al. 2003)  during 
5 years of observations with typical exposures in the range 3000--8500 ksec (see 6th column of table 1); 
these are the same data used to obtain the 4th IBIS survey.
ISGRI images for each available pointing were generated in various energy 
bands using the ISDC offline scientific analysis software OSA (Goldwurm et al. 2003) version 7.0. 
Count rates at the position of the source were extracted from individual images in order to provide light curves 
in various energy bands; from these light curves, average fluxes were then extracted and combined to produce an average source 
spectrum (see Bird et al. 2010 for details). 
In the last column of table 1 the variability indicator as defined by Bird et al. (2010) is also reported.

The \emph{XMM-Newton} data were then fitted together with \emph{INTEGRAL} average spectra   
using {\sc XSPEC} v. 12.5.1 (Arnaud 1996) to cover the broad band from 0.5 to 110 keV.
A detailed description of the results obtained by this  spectral analysis  is given in a dedicated 
section for each source.

\begin{table*}
\begin{center}
\caption{\emph{INTEGRAL}/IBIS position of the sources and locations of the objects detected by \emph{Newton-XMM} within the 90\% 
high energy error circles,  with relative counterparts.  Also the XMM error radii are given at 90\% confidence level.} 
\label{tab2}
\begin{tabular}{lccccc}
\hline
\hline
XMM source  & R.A.     &     Dec   &   error   &  Count rate (0.2--12 keV)  &  Likely counterpart \\ 
  &   (J2000) &  (J2000) &   (arcsec)  & (10$^{-1}$ counts s$^{-2}$)  &        \\
\hline            
\multicolumn{6}{c}{{\bf IGR J15359-5750}  (R.A.= $15^{\rm h}36^{\rm m}00^{\rm s}.0$ (l=323.45), Dec=
$-57^\circ49^{\prime}01^{\prime \prime}.20$ (b=-1.66), error radius = 2$^{\prime}$.2)}\\
   &    &   &   &  &     \\
\#1 (in 90\%) & $15^{\rm h}36^{\rm m}02^{\rm s}.99$ & $-57^\circ48^{\prime}52^{\prime \prime}.2$ & 3.2
&  1.37$\pm$0.04&  2MASS 15360282-5748529\\
& & & & & MGPS J153602-574854 \\
\hline
\multicolumn{6}{c}{{\bf AX J1739.3-2923} (R.A. = $17^{\rm h}39^{\rm m}21^{\rm s}.84$ (l=358.89), Dec =
$-29^\circ23^{\prime}24^{\prime \prime}.00$(b=0.92), error radius = 4$^{\prime}$.5)}\\
   &    &   &   &  &    \\
\#1 (in 90\%) & $17^{\rm h}39^{\rm m}18^{\rm s}.28$ & $-29^\circ23^{\prime}48^{\prime \prime}.7$ & 3.3 &
$0.63\pm0.03$ & 2MASS 17391792-2923478  \\
\hline
\multicolumn{6}{c}{{\bf AX J1740.2-2903} (R.A. = $17^{\rm h}40^{\rm m}11^{\rm s}.1$(l=359.28), Dec  =
$-29^\circ02^{\prime}54^{\prime \prime}.00$ (b=0.95), error radius = 2$^{\prime}$.7)}\\
   &    &   &   &  &   \\
\#1 (in 90\%) & $17^{\rm h}40^{\rm m}18^{\rm s}.01$ & $-29^\circ03^{\prime}37^{\prime \prime}.4$ & 3.3
& $1.54\pm0.06$  &  2MASS 17401814-2903381  \\
\hline
\multicolumn{6}{c}{{\bf IGR J17445-2747} (R.A.  = 17$^{\rm h}44^{\rm m}27^{\rm s}.84$ (l=0.86), Dec =
$-27^\circ45^{\prime}57^{\prime \prime}.6$ (b=0.83), error radius = 2$^{\prime}$.2)}\\
  &    &   &   &  &   \\
\#1 ($\star$) & 17$^{\rm h}44^{\rm m}29^{\rm s}.41$ & -27$^\circ46^{\prime}08^{\prime \prime}.9.3$ & 5.1
& $5.20\pm1.84 $  &  2MASS 17442946-2746114  \\
\hline
\multicolumn{6}{c}{{\bf IGR J18538-0102} (R.A. = $18^{\rm h}53^{\rm m}50^{\rm s}.16$ (l=21.28), Dec  =
$-01^\circ02^{\prime}02^{\prime \prime}.40$ (b=-1.00), error radius = 4$^{\prime}$.8)}\\
   &    &   &   &  &    \\
\#1 (in 90\%) & $18^{\rm h}53^{\rm m}48^{\rm s}.46$ & $-01^\circ02^{\prime}29^{\prime \prime}.7$ & 3.2 &
 $2.62\pm0.08$ &  2MASS 18534847-0102295  \\
&                      & & & & 1RXH J185348.2-010228	  \\
\hline
\hline
\end{tabular}
\end{center}
($\star$):  XMM-Slew source, see text
\end{table*}

\section{Results}

All together we analysed the \emph{XMM} data of 6 IBIS sources (see table 1). 
All sources discussed in this section 
appear in the fourth IBIS catalogue (Bird et al. 2010) either as new detections
(AX J1740.2-2903 and IGR J18538-0102)
or as already known hard X-ray emitters (IGR J15359-5750, 
IGR J17331-2406, AX J1739.3-2923 and IGR J17445-2747).
Apart  from the case of IGR J17331-2406 for which we could not find any XMM counterpart within the IBIS 
positional uncertainty and that of IGR J17445-2747 
where the only XMM detection inside the IBIS error box is possibly not the correct counterpart of the high energy source,
for the remaining  4 sources  we were able to find a convincing counterpart in the X-ray band.
For these sources it was also possible to extract the X-ray spectrum and to perform a broad band analysis.
Table 2  lists the sources with X-ray detections together with their IBIS position and relative uncertainty 
as reported in  Bird et al. (2010). For each of these gamma-ray emitters, we provide the position 
and relative uncertainties (at 90\% confidence level) of 
the sources detected by XMM within the 90\% IBIS error circle and considered to be the 
likely X-ray counterparts (see section on individual sources);
we also report their count rate in the 0.2--12 keV energy band and optical/infrared  identification 
obtained from  various on-line archives such 
as NED (NASA/IPAC Extragalactic Database), HEASARC (High Energy Astrophysics Science Archive Research Center) 
and SIMBAD (Set of Identifications, Measurements, and Bibliography for Astronomical Data). \\
The results of the broad band spectral  analysis relative to the 4 objects having good quality data are reported in table 3
where we list the Galactic absorption, the column density in excess to the Galactic value, the power law photon index, the reduced 
$\chi^{2}$ of the best-fit model together with  the 2--10 keV and the 20-100 keV fluxes.
All quoted errors correspond to 90$\%$ confidence level for a single
parameter of interest ($\Delta\chi^{2}=2.71$).
In the fitting procedure, a multiplicative constant, C, has been introduced to take into account 
possible cross-calibration mis-matches between the X-ray and the soft gamma-ray data.  When treating the INTEGRAL data,	
this constant C	was found to be close to 1  using various source typologies (e.g. 
Masetti et al. 2007; De Rosa et al. 2008; Panessa et al. 2008, Molina et al. 2009), so that any
significant deviation from this value can be confidently ascribed to source flux variability between to non-simultaneous observations.

In the following, results on each individual source are presented.

\begin{figure}
\includegraphics[width=1.0\linewidth]{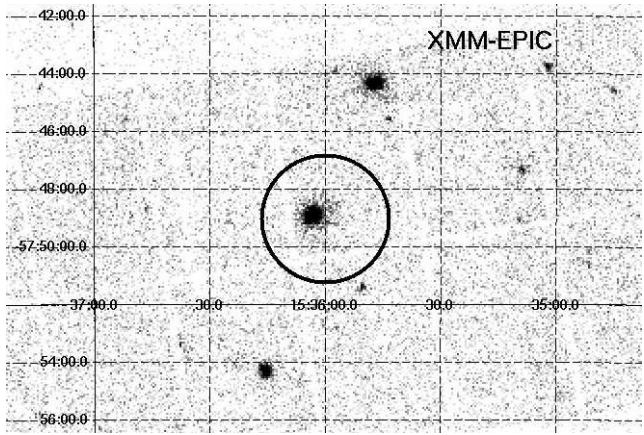}
\caption{EPIC  0.2--12 keV image of the  IGR J15359-5750, only one source is detected by XMM within the IBIS uncertainty  (circle).}
\end{figure}

\begin{figure}
\includegraphics[angle=-90,width=1.0\linewidth]{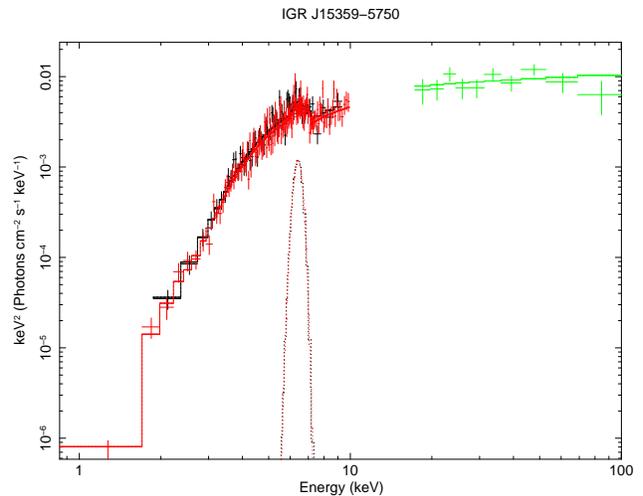}
\caption{ 0.5--110 keV broad band spectrum of  IGR J15359-5750 modeled with a exponentially cut-off power law absorbed by a complex absorption.}
\end{figure}

\begin{figure}
\includegraphics[angle=-90,width=1.0\linewidth]{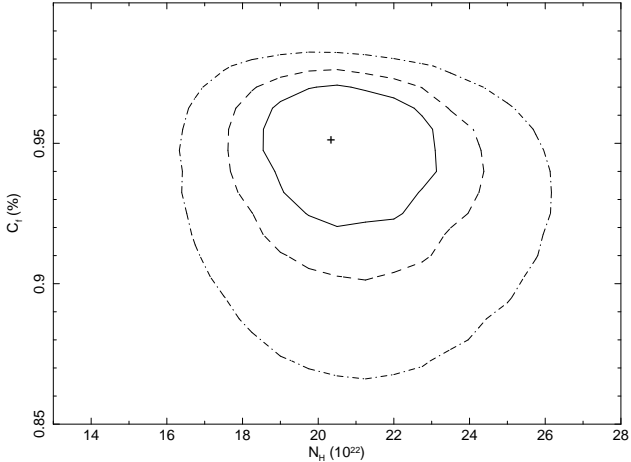}
\caption{Contours 
corresponding to the 68\%, 90\%, and 99\% statistical confidence regions for the two interesting parameters plotted:
column density N$_{H}$ versus its covering fraction C$_{f}$.}
\end{figure}

\subsection{\bf IGR J15359-5750} 
IGR J15359-5750 is the  {\it oldest} known source in our sample, in that  it was reported as a 
new high energy emitter in the second IBIS survey (Bird et al. 2006) but it remains unidentified.
According to an  archival search, a ROSAT Faint X-ray source, 1RXS J153552.8-575055.0,  
is present within the ISGRI error circle with an associated 
positional error of 19$^{\prime \prime}$.
The 0.2-12 keV EPIC image is shown in figure 1 together with the IBIS error circle (2$^{\prime}$.2, 
as reported in the fourth survey) superimposed.  
It is clear that there is only one X-ray source consistent with the high energy position (90\% error circle)
which is not coincident with the ROSAT faint source. 
Instead within  the \emph{XMM-Newton} positional error of 3$^{\prime\prime}$.2 we find an infrared source, 
2MASS 15360282-5748529 which also coincides with the radio object MGPS J153602-574854 having  a 36 cm  flux of 29.4$\pm$1.3 mJy.
The 2MASS object  has  the following magnitudes: $J=14.75\pm0.089$, $H=13.243\pm0.093$ and $K=12.258\pm0.067$ and no optical
counterpart is reported in various archives queried. 
Despite its location on the Galactic plane,  the presence of 
radio and X-ray emission  strongly suggest that this source is a background AGN.\\
The combined \emph{XMM-Newton} and \emph{INTEGRAL}-IBIS spectrum supports this indication.
A model consisting of a simple power law absorbed by only  Galactic absorption   provides a poor fit to the data; furthermore  
the  photon index obtained is too  flat ($\Gamma=-0.98$) and the XMM/IBIS cross calibration constant (C) much lower than 1;
this is probably an indication for  the presence of more absorption, this time intrinsic to the source. 
The addition of intrinsic  column density  is strongly required by the data ($\Delta \chi^{2}$ = 846 for 1 d.o.f.): its value 
is found to be of N$_{H}$ = 2 $\times$ 10$ ^{23}$ cm$^{-2}$ 
and the photon index has in this case a more canonical value of $\Gamma \sim 1.7$ and also C is $\sim$0.95. Despite the fit improvement, 
the data to model ratios show excess emission at soft X-ray energies and evidence for a line  at around 6 keV.
Another significant improvement ($\Delta \chi^{2}$=32 for 2 d.o.f.) is indeed  obtained when we introduce the k$_{\alpha}$ 
iron fluorescence emission line at a fixed energy of 6.4 keV: the line is broad with a $\sigma$ of  0.34$^{+0.31}_{-0.13}$ keV  
and an equivalent width  of 327$^{+195}_{-125}$  keV. 
To account for the soft excess, we tried a more complex absorption model where besides  the intrinsic column 
density, another absorber partially covering the source (model \texttt{pcfabs} in  \texttt{XSPEC}) is also considered.  
The addition of this extra absorption component   provides a good fit ($\chi^{2}$ =223/206) and gives a photon index 
of $\Gamma$ = 1.85  and column densities of the order of N$_{H_{1}} \sim 4 \times 10^{22} cm^{-2}$ fully covering the source  
and N$_{H_{2}} \sim 2 \times 10^{23} cm^{-2}$
covering only   0.95\% of the object. In figure 2 and table 3 the unfolded broad band spectrum fitted with this model is
shown and described, while in figure 3 the contours relative to the complex absorber (N$_{H_{2}}$ versus its covering fraction) are displayed.
The cross-calibration constant between XMM and INTEGRAL has a value of C = 1.1$^{+0.51}_{-0.25}$ i.e. consistent with unity.
Next we fixed the cross-calibration constant to 1 and
 substituted the simple power law with an exponentially cut-off power law spectrum reflected by neutral material (model \texttt{pexrav} 
in  \texttt{XSPEC}): in this way we were able to provide constraints on the reflection  component  (R$<$1.65)  and  cut-off energy  
(E$_{C}$ $\ge$ 68 keV) .\\
The overall spectral characteristics  are clearly compatible with the suggested extragalactic nature of the source 
and further indicates that IGR J15359-5750 could be an intermediate  type  AGN like Mrk 6, 4U 1344-60 or IGR J21247+5058 (see Molina et al. 2009).

\subsection{\bf IGR J17331-2406}   
IGR J17331-2406, first  reported as a high energy emitter by Krivonos et al. (2007), is listed in the 4th IBIS survey catalog as a transient source
since to date it was detected by INTEGRAL only once;  it was discovered on September 2004 during outburst activity 
lasting $\sim$ 30 days and reaching  a peak flux of $\sim$ 30 mCrab or 4$\times$10$^{-10}$ erg cm$^{-2}$ s$^{-1}$ (18--60 keV). 
The spectrum was quite hard, fitted with a power law having $\Gamma$ $\sim$ 1.8 (Lutovinov et al. 2004). 
In the \emph{XMM-Newton} observation reported in this work   we could not find any X-ray source 
within the ISGRI error circle; the X-ray upper limit to the flux is equal to 
$\sim$ 2$\times$10$^{-14}$ erg cm$^{-2}$ s$^{-1}$ (0.5--10 keV).
It is worth noting that other 0.5--10 keV observations of the source have been reported in the literature from both Swift/XRT and Chandra 
(Landi et al. 2010, Thomsick et al. 2008), but no X-ray detections were obtained, providing upper limits 
very similar to that obtained with XMM. 
If we extrapolate the spectrum detected by IBIS during the outburst to low energies we obtain a flux of $\sim$6$\times$10$^{-11}$ erg cm$^{-2}$ s$^{-1}$  
in the 0.5-10 keV band implying a dynamical range from quiescent to outburst of the order of 3000.
It is clear that it will be extremely difficult to find the soft X-ray counterpart of IGR J17331-2406, unless a  monitoring program is planned 
with the aim of promptly observing the source when in outburst.

\begin{figure}
\includegraphics[width=1.0\linewidth]{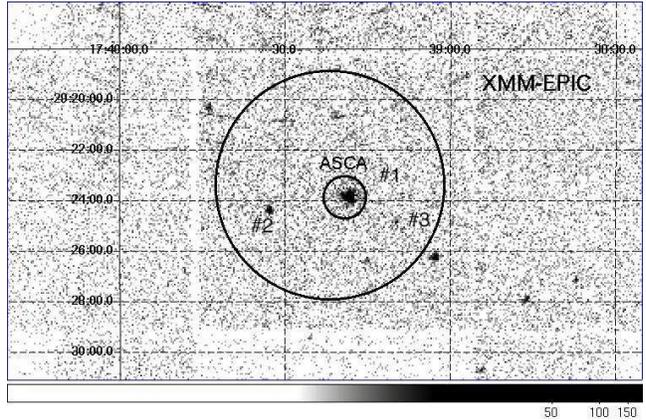}
\caption{EPIC 0.2--12 keV image of the region surrounding AX J1739-2923. 
Also plotted are the 90\% ASCA (small) and IBIS (larger)  error circles.}
\end{figure}

 \begin{figure}
\includegraphics[angle=-90,width=1.0\linewidth]{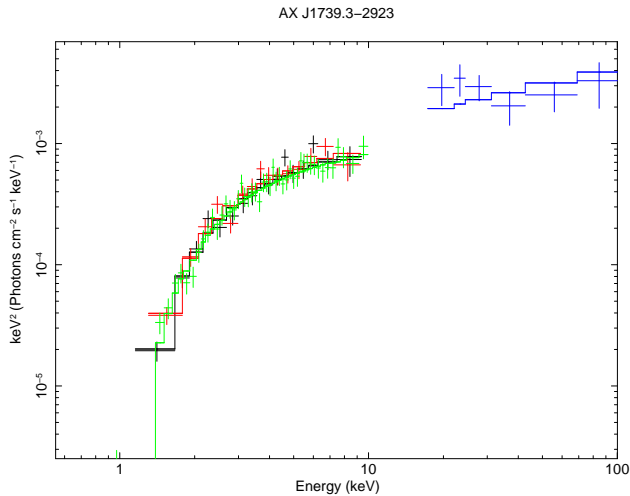}
\caption{XMM-Newton and INTEGRAL-ISGRI broad band spectrum of AX J1739.2-2923 modeled with a power law absorbed by the Galactic plus an 
intrinsic absorption.}
\end{figure}

\subsection{\bf AX J1739.3-2923}     
AX J1739.3-2923 was first  reported as a high energy emitter in the 3rd IBIS survey (Bird et al. 2007)
and soon associated to an ASCA object detected   during observations of the  Galactic center region  
(Sakano et al. 2002).\\
The ASCA and  the most recent forth IBIS error circles  (50$^{\prime \prime}$ and  4$^{\prime}$.5 at 90\% respectively) are shown in figure 4, 
superimposed on the  0.2-12 keV  EPIC image. \\
It is  evident from this figure that there are two bright X-ray sources  and one weak source located inside the IBIS error circle: source N. 1 is also compatible with
the ASCA positional uncertainty and is clearly AX J1739.3-2923 while the other two are much weaker  X-ray emitters and are
detected by XMM for the first time. Besides being the brightest of the two, the ASCA  source  is also the harder as 
it  remains visible above 4.5 keV while the other disappears above a few keV. Thus the observational evidence
suggests that AX J1739.3-2923 is the true  counterpart of the INTEGRAL object but it is now  located with arsec precision thanks to XMM. 
Only marginally compatible with the   XMM error circle  we find one possible counterpart in a
2MASS pointed source 2MASS 17391792-2923478 located at 4.8$^{\prime\prime}$
from AX J1739.3-2923  having 
magnitudes  $J\sim$12.96, $H=12.17\pm0.053$ and $K=11.65\pm0.052$; here too no optical counterpart was found.\\
Next we combined the  \emph{XMM-Newton} MOS and pn data together with the \emph{INTEGRAL}-IBIS points  in order 
to study the broad band (0.5-110 keV) spectrum of the source.
As a first attempt, we tried a fit with a simple power law absorbed by the Galactic column density which in the direction of the source 
is quite high (9.8 $\times$10$^{21}$ cm$^{-2}$,  Dickey \& Lockman  1990) due to its  location on the Galactic plane. 
This simple model fits the data poorly ($\chi^{2}$=305/166)  giving a very hard photon index ($\Gamma$=0.68) which 
suggests the presence of  extra absorption intrinsic to the source. Indeed the addition of this extra component is highly required by the data  ($\Delta\chi^{2}$=149 for 1 d.o.f);
this fit gives  a column density  of  N$_{H}$=1.88  $\times$10$^{22}$ cm$^{-2}$,  a photon index $\Gamma$=1.52
and a 2-10 keV observed flux of 1.3 $\times$ 10$^{-12}$ erg cm$^{-2}$ s$^{-1}$.
The unfolded broad band spectrum using this model is shown in figure 5 (see also table 3).
Our spectral parameters are in good agreement with the ASCA results (Sakano et al. 2002), in particular the very similar flux values   
suggest that the source is persistent and not remarkably variable; also the cross-calibration constant between XMM and INTEGRAL 
is C=1.59$^{+0.72}_{-0.51}$ i.e. marginally compatible within uncertainties with unity. 
An inspection  of the IBIS long term light curve  (18--60 keV), spanning about 4.5 years, shows no sign of significant variability; this 
confirms the previous hypothesis of a persistent nature for the source. 
No radio counterpart, generally expected for an extragalactic object, 
has been found using all available radio catalogs in the HEASARC database. 
On the other hand, the location of the source 
on the Galactic plane, the high intrinsic absorption, the hard photon index and the lack of timing signatures such as X-ray 
bursts in spite of the long monitoring, all favour the hypothesis that AX J1739.3-2923  is a weak, persistent and absorbed HMXB system.

\begin{figure}
\includegraphics[width=1.0\linewidth]{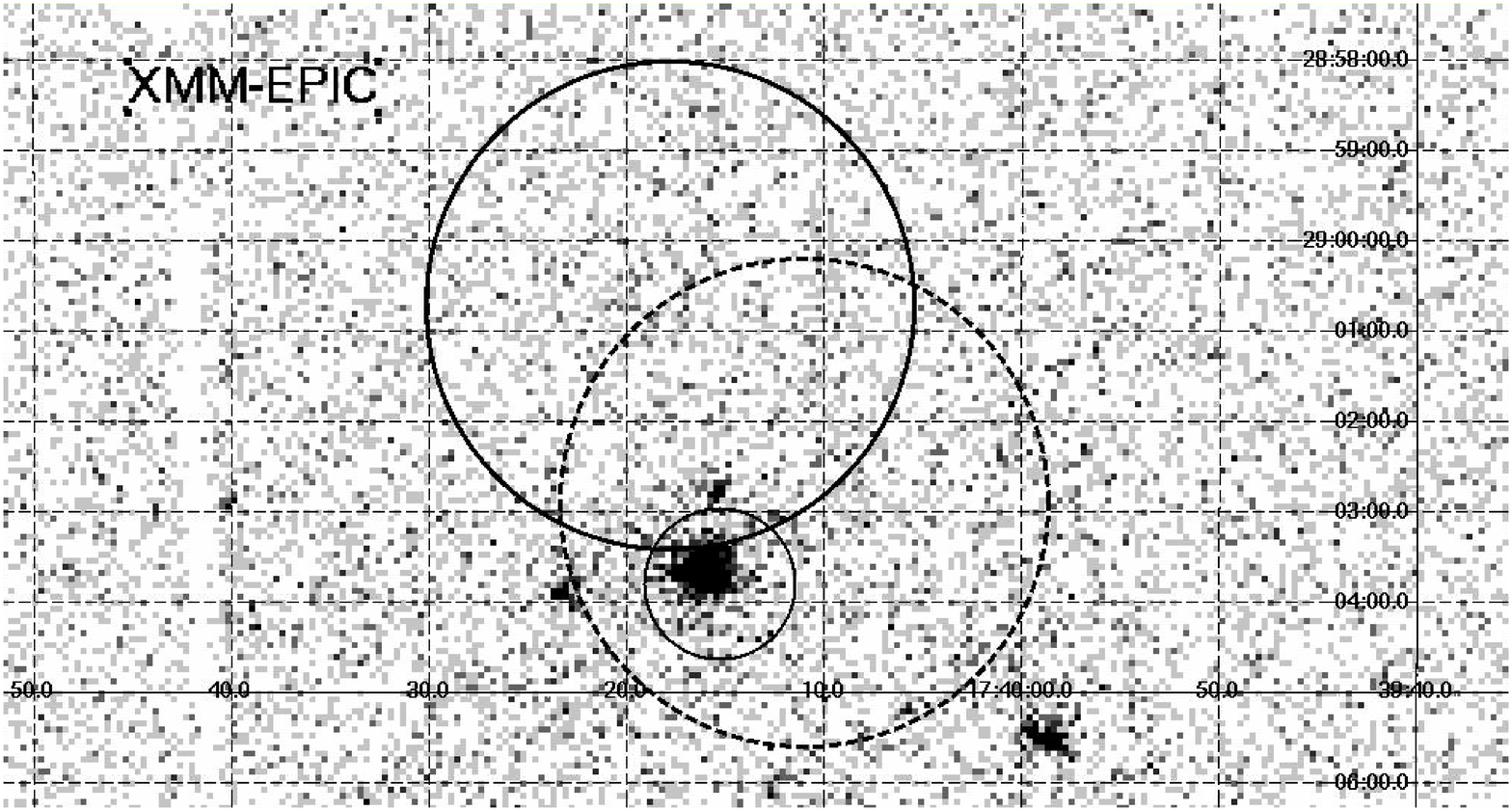}
\caption{EPIC 0.2--12 keV image of the region surrounding AX J1740.2-2903.
Superimposed are the  4th IBIS catalogue  (solid large), IBIS re-defined (dashed large) and ASCA (small) error circles.}
\end{figure}

\begin{figure}
\includegraphics[angle=-90,width=1.0\linewidth]{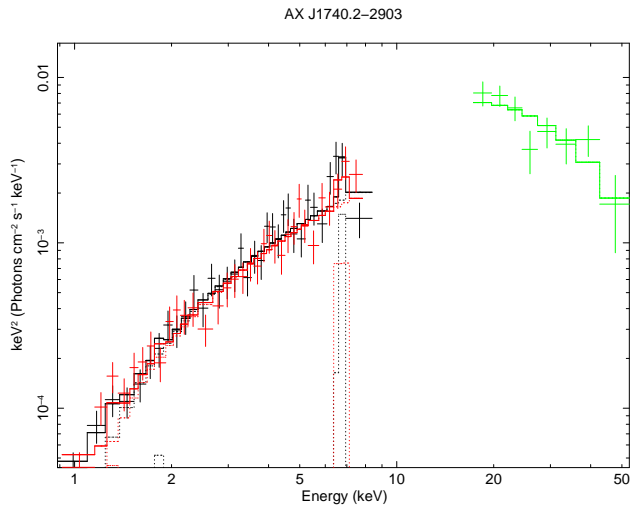}
\caption{Broad band (0.5--60 keV) spectrum of AX J1740.2-2903 fitted with a Raymond-Smith plasma
plus a cut-off power law and an iron line.}
\end{figure}

\subsection{\bf AX J1740.2-2903} 
AX J1740.2-2903 is an X-ray source discovered by ASCA 
during the Galactic center region scans (Sakano et al. 2002). It is listed in the latest 
IBIS survey catalog (Bird et al. 2010) with a 20--40 keV flux of 0.5 mCrab or 3.8$\times$10$^{-12}$ erg cm$^{-2}$ s$^{-1}$
and  it  was not reported in the previous 3rd IBIS catalog because its detection (5.5$\sigma$) was just below the significance threshold 
considered in a region, the Galactic Centre, where large systematics errors  are present.
Investigation of  the  IBIS long term light curve of AX J1740.2-2903 which 
spans about 4 years,  shows no sign of flaring or transient emission, instead 
the source seems to show a rather weak persistent emission.\\
The error circles of both ASCA (50$^{\prime \prime}$ at 90\%) and IBIS   (2$^{\prime}$.7  at 90\% solid circle) 
have been superimposed on the  0.2-12 keV  EPIC image in figure 6 which reveals that  AX J1740.2-2923 
is just at the border of INTEGRAL positional uncertainty.  
Due to location of  AX J1740.2-2923  being near the Galactic Centre where there are some systematic 
structures in the ISGRI sky maps, we have re-analysed the data and got an improved position
at R.A.(J2000) = $17^{\rm h}40^{\rm m}11^{\rm s}.1$, Dec(J2000) =$-29^\circ02^{\prime}54^{\prime \prime}.00$
and using these coordinates the X-ray source is found to be well inside the high energy error circle (dashed circle in figure 6).
Furthermore, considering that AX J1740.2-2923 
is the only X-ray source in the XMM field detected above 4.5 keV,  we conclude that
the ASCA source is the natural IBIS counterpart, now  located by XMM with higher precision.\\
We found only  an  infrared source, 2MASS 17401814-2903381,  within the  XMM positional uncertainty:
it has J, H and K magnitudes  of 13.34, 11.52 and 10.52 respectively and no optical and radio counterpart.\\
Despite the lack of EPIC pn data, we were nevertheless capable of performing the  analysis 
of the broad band spectrum of the source.
A fit with a power law absorbed by the Galactic column density does not provide a good fit ($\chi^{2}$=156/71) and
gives a hard photon index of about 1.  
Contrary to the previous cases, the addition of an extra absorption component is not required  by the data 
and, furthermore we found a very low value (0.43) for  the cross calibration constant between XMM and INTEGRAL.
We note   that the low (0.5-10 keV) and high (17-110 keV) energy 
data sets  have completely different spectral shape:  while  the X-ray spectrum  is well fitted with a power law of photon index 
$<$1, the IBIS spectrum is very soft being fitted with a power law having $\Gamma$=3.2. 
Matching  these two spectra clearly requires the  presence of  a low energy cut-off if we assume no variability of the source, or  a change in spectral shape if we suppose that the flux can vary. 
Fitting  the source broad band (0.5-110 keV) spectrum of AX J1740.2-2903 with
an exponentially cut-off power law (model \texttt{cutoffpl} in  \texttt{XSPEC}) absorbed by the Galactic column density, 
provides an acceptable  fit ($\chi^{2}$=93/70), a  hard photon index  $\Gamma$ = 0.6$^{+0.20}_{-0.14}$ and a cut-off energy at E$_{C}$=11$^{+3}_{-2}$.
The XMM/IBIS constant is 2.72$^{+1.3}_{-0.9}$ not consistent with unity which indicates some
variability between the pointed XMM observation and the average IBIS measurement.
However, we must note that
extracting a spectrum without possible contamination from nearby sources in such a crowded region as the Galactic Centre, is difficult
so this might also explain the non-ideal cross-calibration constant.
Inspection of the data to model ratio suggests the presence of residuals in the soft part of the spectrum around  
6-7 keV. 
Therefore we first tried to fit the soft-excess both with a black body and thermal plasma models but the best fit 
only improves by adding to the cutoff power law a Raymond-Smith
plasma ($\chi^{2}$=73/68). The temperature of the plasma is found to lie in the range 0.16-1.04  keV. \\
The addition of a line left free to vary in the 6-7 keV energy range, also  improves the fit (but only at the  95\% confidence level) 
and provides  our best fit model presented in figure 7 (see also table 3).
Fitting the data with this model gives a flatter photon index $\Gamma \sim$0.4, a cut-off energy at around 10 keV,
a line at 6.7$^{+0.12}_{-0.40}$ keV with  EW of 400$^{+400}_{-300}$ eV.
In the light of all above findings, here too we suggest for this source a likely persistent HMXB nature.

\begin{figure}
\includegraphics[width=1.0\linewidth]{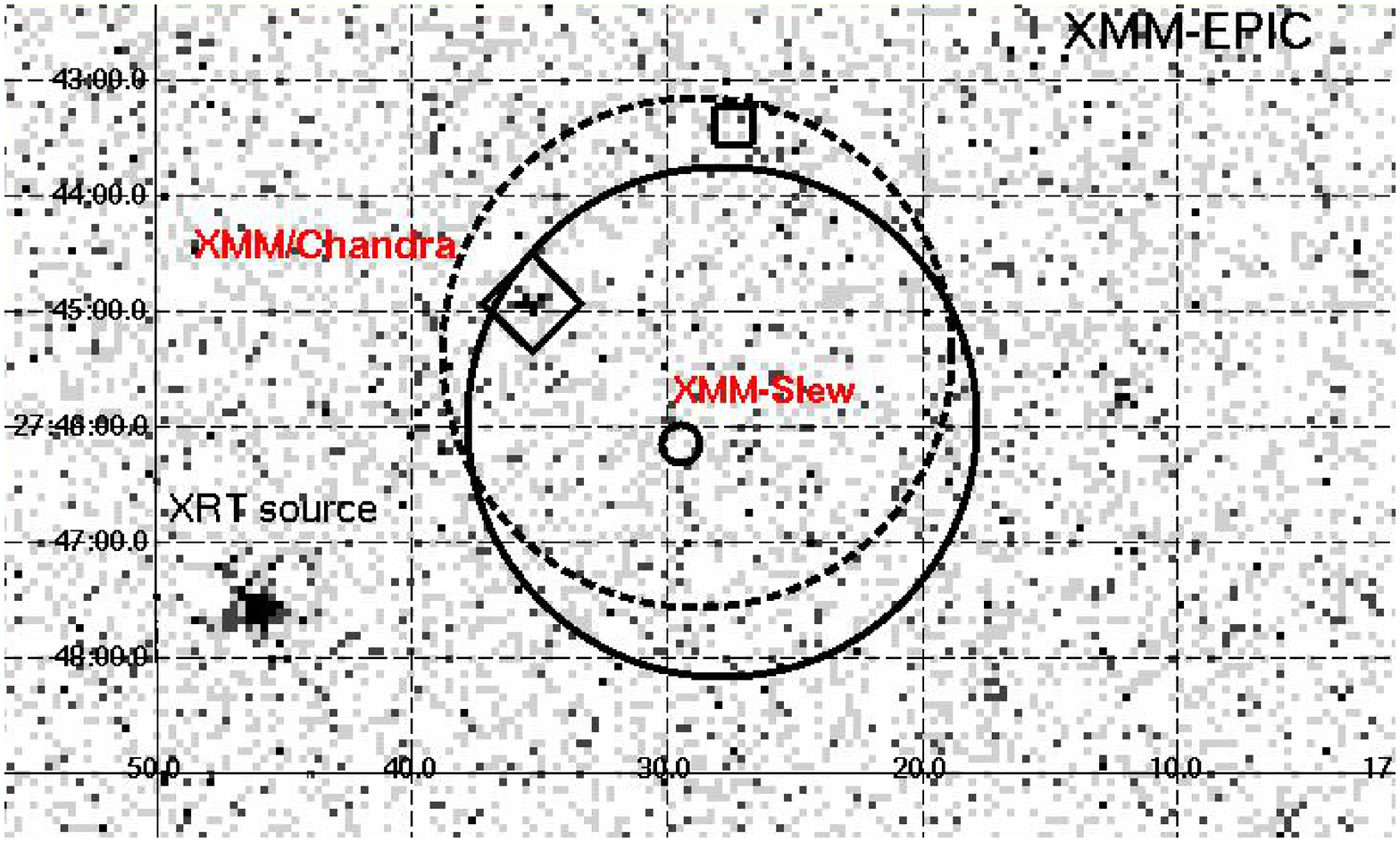}
\caption{0.2-12 keV XMM-EPIC image of the region of IGR J17445--2747. All the possible X-ray counterparts proposed so far are also reported
together with the IBIS error circles (dashed from the 3rd catalogue and continued from the 4th).}
\end{figure}

\subsection{\bf IGR J17445-2747} 
IGR J17445-2747 represents an interesting case of a source with many X-ray follow up observations but still without a convincing counterpart.
Figure 8 shows the \emph{Newton-XMM} 0.2-12 keV image with superimposed all the possible X-ray counterparts proposed so far together with the 
\emph{INTEGRAL-IBIS} error circles (dashed from the 3rd catalogue and continuous from the 4th).
IGR J17445-2747 first appeared in the 3rd IBIS survey (Bird et al. 2007) and was subsequently associated to a relatively faint Swift/XRT source  
with a 2-10 keV flux of 1.5 $\times$ 10$^{-13}$ erg cm$^{-2}$ s$^{-1}$,   located however outside  the INTEGRAL error box 
(Landi et al. 2007).\footnote{Note that the source was wrongly put inside the IBIS error box in the Atel.} \\
Then Chandra follow up observations of the region detected three sources (Tomsick et al. 2008): the same source detected by XRT, a new  one located 
inside the new ISGRI error circle (J174435.4-274453, diamond point)  reported in the 4th IBIS catalogue and a third one
 just at its border (J174427.3-274324 box point).\\
Within the INTEGRAL/IBIS error circle 
\emph{Newton-XMM} detects  only one source at a position compatible with that of the Chandra source J174435.4-274453
indicating that this could be a possible counterpart. 
The source spectrum is poorly sampled by XMM as this is only a 4.6 sigma detection 
but we were able to estimate the  0.2-12 keV observed flux of 8 $\times$ 10$^{-14}$ erg cm$^{-2}$ s$^{-1}$ fully 
compatible with that of Chandra of $\sim$ 7 $\times$10$^{-14}$ erg cm$^{-2}$ s$^{-1}$.\\
However, we note that  within the IBIS error box there is also an XMM-Slew source: XMMSL1 J174429.4-274609 (circle point in the figure) 
whose coordinates are reported in table 2. This source with a 0.2-12 keV flux of 1.64 $\times$ 10$^{-12}$ erg cm$^{-2}$ s$^{-1}$
is the brightest in the high energy error circle and it is also extremely variable as it was seen only once out of four observations of the region
made at different epochs.
This source is associated only to an infrared object (2MASS 17442946-2746114) within the  XMM-Slew (5.1$^{\prime\prime}$) positional uncertainty
with J, H and K magnitudes of  15.13, 12.88 and 12.85 respectively.\\
The XMM upper limit on the source flux is 0.4 $\times$ 10$^{-13}$ erg cm$^{-2}$ s$^{-1}$ in the same waveband 
implying a dynamic range of around 40.\\
IGR J17445-2747 is reported   in the 4th IBIS catalogue as a transient bursting source since it was significantly detected 
at $\sim$ 13$\sigma$ level (20--100 keV) only during its outburst activity lasting for a total of $\sim$ 30 days and reaching a peak flux 
of $\sim$ 30 mCrab or 4.6$\times$ 10$^{-10}$ erg cm$^{-2}$ s$^{-1}$ (20--100 keV). On the contrary the source 
was not detected in the total dataset for an on-source exposure time of $\sim$ 7.3 Ms, providing an upper 
limit to the flux of 0.1 mCrab (20--40 keV) and resulting in a dynamical range of $\sim$ 300. \\
Therefore, given the transient nature of both IGR J17445-2747 and XMMSL1 J174429.4-274609, 
we conclude that the two sources are very likely associated.

\begin{figure}
\includegraphics[width=1.0\linewidth]{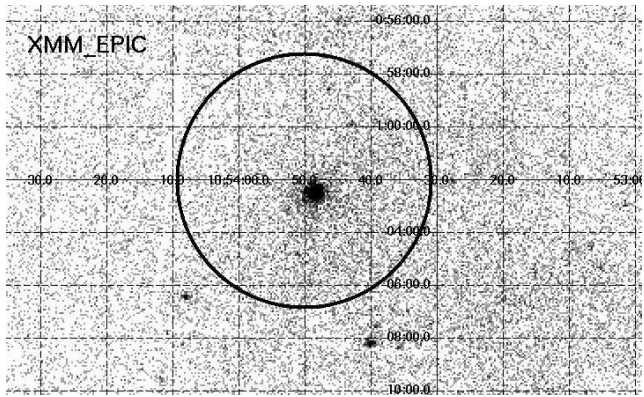}
\caption{XMM-EPIC 0.2--12 keV image of IGR J18538-0102 with the high energy error box superimposed.}
\end{figure}

\begin{figure}
\includegraphics[angle=-90,width=1.0\linewidth]{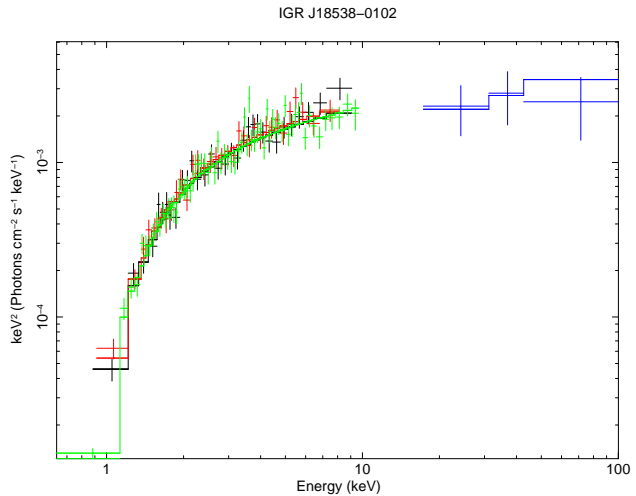}
\caption{XMM-EPIC and INTEGRAL IBIS (0.5-110 keV) spectrum of IGR J18538-0102 fitted with an absorbed power law.}
\end{figure}

\subsection{\bf IGR J18538-0102} 
IGR J18538-0102 is a newly discovered INTEGRAL source listed in the fourth IBIS 
survey (Bird et al. 2010). 
Recently Stephen et al (2010) provided  an improved
position for the source using the association with an XMM Slew catalogue source (2XMM J185348.4-010229). 
These authors also noted that  IGR J18538-0102 is spatially coincident with a hot spot in the supernova remnant candidate G32.1-0.9 detected 
in X-rays by ROSAT and ASCA (Folgheraiter et al. 1997):
 but it has a much harder spectrum and higher absorption than observed in the supernova remnant. 
This suggests the possibility that IGR J18538-0102 could be a more distant Galactic source or 
a background AGN and its alignment with G32.1-0.9 is only coincidental.  Halpern and Gotthelf (2010) then
reported on the XMM
observation used here and further discussed a possible infrared/optical counterpart of the source.
We re-analysed the XMM data in order to combine them with IBIS ones, performing  for the first time a broad band  spectral analysis and discussing
further the nature of this object.
The EPIC 0.2-12 keV image shown in figure 9 with the INTEGRAL 90\% error circle indicates 
that there is  only one X-ray counterpart. 
Archival searches within the \emph{XMM-Newton} positional uncertainty  finds the infrared object 
(2MASS 18534847-0102295) discussed by Halpern and Gotthelf which has J, H, and K magnitudes of 14.16, 14 and 12.50 respectively.
It coincides with the USNO-B1 0889-0406090 object (R= 15.2 magnitudes) and with the soft X-ray source 1RXH J185348.2-010228
detected by the ROSAT-HRI.\\
Next, we concentrated on broad band spectral analysis.
Folgheraiter and co-workers (1997) already provided indications on the source X-ray shape: they 
found that either a power law with a photon index of  1.8 or, alternatively, a thermal model with a kT of about 2 keV
was appropriate to fit the 0.5--2 keV low energy X-ray (ROSAT and ASCA) data. 
Both models required absorption in excess to the Galactic value which in the source direction
is 9.93 $\times$ 10$^{21}$ cm$^{-2}$  (Dickey \& Lockman, 1990).
Our  broad band 0.5-110 keV  spectrum is poorly fitted with a simple power law absorbed by the Galactic 
absorption ($\chi^{2}$=307/260) but the addition of extra  absorption, probably intrinsic to the source improves 
the fit significantly (at a confidence level $>$ 99.99\%) and provides  the best description of the source spectrum ($\chi^{2}$=255/259).
The amount of extra column density is  N$_{H}$ $\sim$4 $\times$ 10$^{21}$ cm$^{-2}$ while 
 the value of photon index is $\Gamma$=1.56; the cross calibration is found to be 0.65$^{+0.28}_{-0.22}$
indicating  possible  variability between the pointed XMM observation and the IBIS average measurement.  
The combined XMM/INTEGRAL unfolded spectrum fitted with this model is shown in figure 10 and described in table 3.
The spectral parameters are quite in agreement with what found by Halpern from the XMM data analysis 
alone but not fully compatible with the spectral parameters  reported by Folgheraiter et al. (1997) although we get
a similar flux in the 0.5-2 keV energy band ($\sim$2  $\times$ 10$^{-12}$ cm$^{-2}$ s$^{-1}$).
On the other hand, it is worth noting that a thermal model does not fit our broad band data.\\
We conclude that IGR J18538-0102, as already argued by Halpern \& Gotthelf (2010), 
is unlikely a compact object in the supernova remnant G32.1-09
but could be a background AGN that is coincidentally  aligned  with the supernova.

\begin{table*}
\begin{center}
\caption{\emph{XMM-Newton} and  \emph{INTEGRAL} broad band spectral analysis results: best fit results. }
\label{tab3}
\begin{tabular}{lcccccc}
\hline
\hline
 Source & $N_{\rm H}(Gal)$ & $N_{\rm H}$ & $\Gamma$ & $\chi^2/\nu$  & $F^{(\star)}_{\rm (2-10~keV)}$ & $F_{\rm (20-100~keV)}$\\
  & ($10^{22}$ cm$^{-2}$) &  ($10^{22}$ cm$^{-2}$)   &  &  &   ($10^{-12}$ erg cm$^{-2}$ s$^{-1}$) & ($10^{-12}$ erg cm$^{-2}$ s$^{-1}$)  \\
\hline
\hline
IGR J15359-5750  &  1.190   & 20.1$^{+2.5}_{-2.9}$ ($\dagger$) (0.95$^{+0.02}_{-0.04}$) & $1.85^{+0.27}_{-0.25}$ & 223/207  &  4.97  &  23.28 \\
                                              &                & 4.26$^{+0.98}_{-0.75}$  &                                       &                  &                           \\ 
\hline
AX J1739.3-2923              & 0.980     & 1.88$^{+0.35}_{-0.31}$    & $1.52^{+0.14}_{-0.14}$ & 155/165       & 1.27  & 6.4 \\
\hline
AX J1740.2-2903              & 0.968     &             -                           &  0.36$^{+0.21}_{-0.18}$ &   73/68  &   3.18  & 7.00 \\
\hline
IGR J18538-0102              & 0.993    & 0.41$^{+0.1}_{-0.1}$  &  $1.56^{+0.08}_{-0.08}$ & 255/259 & 4.00 & 7.83 \\   
\hline
\hline
\end{tabular}
\begin{list}{}{}
($^{\star}$) Observed fluxes
($\dagger$) partial covering absorption with its relative covering fraction.\\

\end{list}
\end{center}
\end{table*}

\begin{table*}
\begin{center}
\caption{Summary }
\label{tab4}
\begin{tabular}{ll}
\hline
\hline
 Source &  Possible Identification\\
\hline
\hline
IGR J15359-5750             & AGN of intermediate type \\
IGR J17331-2406             & Black Hole in the Galactic bulge\\
AX J1739.3-2923              &  HMXB \\
AX J1740.2-2903              &  HMXB \\
IGR J17445-2747             &  Galactic transient object \\
IGR J18538-0102             & likely background AGN\\   
\hline
\hline
\end{tabular}
\end{center}
\end{table*}

\section{Conclusions}
In this work we have cross-correlated the list of the still unidentified hard X-ray emitters listed in the 4th IBIS survey 
with the archive of all \emph{XMM-Newton} pointings finding a set of 6 objects with archival data.
First, we studied the EPIC images in order to find in the IBIS error circle the X-ray counterpart(s). 
In the case where an associated source has been found, the \emph{XMM-Newton} data have then been used together the 
\emph{INTEGRAL-IBIS} spectra  to study the broad band slope and investigate the 
possible nature of the source. \\
In table 4 a summary of our proposed identifications is reported.\\
In a couple of cases no obvious X-ray counterpart has been found from the \emph{XMM-Newton} observations, like
IGR J173331--2406 and  IGR J17445-2747. In the first case, no X-ray source has been detected. This is in perfect agreement 
with the IBIS survey data where this source has been found to be transient. 
Extrapolating to the low energies (0.5-10 keV) the spectrum seen 
by IBIS during the source outburst and comparing it with the XMM upper limit, we found a dynamical range of the order of 
3000. Such a high value strongly suggests that IGR J17331--2406 could be either a  transient black hole  in the Galactic bulge
or a SFXT although this latter interpretation can be ruled out due to source  location off the Galactic plane ($\sim$ 5 degrees) 
as well as its significantly longer outburst duration compared to classical SFXTs. \\
The other case where the XMM observation does not provide a secure X-ray counterpart is that of IGR J17445-2747.
From the imaging analysis  we found a faint XMM source at the border of the IBIS error circle 
which has also been detected by both Swift-XRT and Chandra.  On the other hand,  from archival searches we found an XMM slew source well inside the 
high energy positional uncertainty which has been seen only once by XMM  and therefore is again in perfect agreement with the high energy
survey data which classified this source extremely variable; we therefore  associate IGR J17445-2747
to  XMMSL1 J174429.4-274609 .\\
In the remaining four cases we have found a convincing X-ray counterpart in the IBIS error circle for which it has been possible to
search for counterparts in other wavelength bands and also perform the spectral  data analysis in the 0.5--110 keV band.
The spectral parameters obtained together with the possible IR/optical/radio counterpart found allowed us to investigate
on the nature of each source.
We conclude that IGR J15359-5750 is an AGN of intermediate type for which we were able to estimate  the amount of the complex absorption
as well as to give constraints on the reflection and the high energy  cut-off.
For the two ASCA sources AX J1739.3-2923 and AX J1740.2-2903 we suggest a strongly absorbed Galactic nature and for both  we argue that they are likely 
persistent HMXB systems.
More uncertain is the case of IGR J18538-0102 which is spatially coincident with a hot spot in the supernova remnant G32.1-0.9
detected previously by ROSAT and ASCA. From the broad band spectral analysis performed in this work we can conclude that this object is
unrelated compact object  which happen to coincide with the supernova remnant, probably  a background AGN that is coincidentally aligned even if no radio counterpart has been found 
in the more precise XMM error box.\\

\section*{Acknowledgments}
This research has made use of data obtained from the SIMBAD database operated at CDS, Strasbourg, France; 
the High Energy Astrophysics Science Archive Research Center (HEASARC), provided by NASA's Goddard Space 
Flight Center NASA/IPAC Extragalactic Database (NED). We thank the anonymous referee for 
the very detailed and careful review of this paper.
The authors acknowledge the ASI financial support via ASI--INAF grant I/008/07/0.

\end{document}